\documentclass[apj]{emulateapj}
\usepackage[colorlinks = true, linkcolor = blue, urlcolor  = blue, citecolor = blue, anchorcolor = blue]{hyperref}
\usepackage{hyperref}
\newcommand\fnurl[2]{\href{#2}{#1}\footnote{\url{#2}}}

\shorttitle{Active longitude and solar flare occurrences}
\shortauthors{N. Gyenge et al.}

\begin{document}

\title{Active longitude and solar flare occurrences}

\author{N. Gyenge\altaffilmark{1,2} , A. Ludm\'any\altaffilmark{1}, T. Baranyi\altaffilmark{1}}
\thanks{\altaffilmark{}e-mail: gyenge.norbert@csfk.mta.hu}

\affil{\altaffilmark{1}Debrecen Heliophysical Observatory (DHO), Konkoly Observatory,  Research Centre for Astronomy and Earth Sciences, \\ Hungarian Academy of Sciences,
Debrecen, P.O.Box 30, H-4010, Hungary\\
\altaffilmark{2}Solar Physics and Space Plasmas Research Centre (SP2RC), School of Mathematics and Statistics, University of Sheffield\\
Hounsfield Road, Hicks Building, Sheffield S3 7RH, UK\\}

\begin{abstract}
The aim of the present work is to specify the spatio-temporal characteristics of flare activity observed by the Reuven Ramaty High Energy Solar Spectroscopic Imager (RHESSI) and Geostationary Operational Environmental Satellite (GOES) satellites in connection with the behaviour of the longitudinal domain of enhanced sunspot activity known as active longitude (AL). By using our method developed for this purpose, we identified the AL in every Carrington Rotation provided by the Debrecen Photoheliographic Data (DPD). The spatial probability of flare occurrence has been estimated depending on the longitudinal distance from AL in the northern and southern hemispheres separately. We have found that more than the 60\% of the RHESSI and GOES flares is located within $\pm 36^{\circ}$ from the active longitude. Hence, the most flare-productive active regions tend to be located in or close to the active longitudinal belt. This observed feature may allow predicting the geo-effective position of the domain of enhanced flaring probability. Furthermore, we studied the temporal properties of flare occurrence near the active longitude and several significant fluctuations were found. More precisely, the results of the method are the following fluctuations: $0.8$ years, $1.3$ years and $1.8$ years. These temporal and spatial properties of the solar flare occurrence within the active longitudinal belts could provide us enhanced solar flare forecasting opportunity.
\end{abstract}

\section{Introduction}

The spatial distribution and temporal behaviour of flare occurrences are important elements of space weather. Regarding spatial properties, the latitudinal distribution (i.e the butterfly diagram) is easily detectable because it is connected to the well-known varying latitude of the activity during the solar cycle. The determination of the longitudinal distribution is a harder challenge. It was mentioned as early as in the 19th century by   \citet{Carrington1863} that sunspot groups may prefer some longitudinal domains. This was also observed later by \citet{Maunder1905} and \citet{Losh}. In the following decades, it was usually considered that sunspot emergence is not equally probable at all heliographic longitudes but the results of localisation attempts were not always convincing. The causes of the differing results are diverse. 

Two approaches were followed. One of them considered the locations of enhanced activity to be "active nests" as isolated entities of the activity in the works of \citet{Becker55}, \citet{Brouwer1990} and \citet{Castenmiller86}. The other approach addresses active longitudes (hereafter AL) as persistent domains of activity. The presumptions of the applied methods are different and they may have considerable impact on the results. Possible presumptions are the followings: rigidly rotating frame carrying an active longitudinal domain as reported by \citet{Balthasar83}, \citet{Kitchatinov05}, and \citet{Ivanov07}; persistent active longitudes under the influence of differential rotation presented e.g. by \citet{Usoskin05} and \citet{Zhang2011a}; two active longitudes at a distance of $180^{\circ} $ as in \citet{Usoskin05}. There were also sceptic views; \citet{Pelt05} showed that some results may be artefacts of the applied methods, \citet{Henney05} even questioned the existence of active longitudes.

The preferred longitudinal domains of flares have been studied with similar presumptions and methods as those of the sunspot studies. The active nests are called 'superactive regions' \citep{Bai87} or 'hot spots' \citep{Bai88}. The idea of rigid rotation also arises \citep{Warwick65} but the obtained rotation rates are diverse \citep{Bai03a}. The role of differential rotation is also considered and the presumption of the twin active longitudes at $180^{\circ} $ apart \citep{Zhang07, Zhang2011b}. It seems reasonable to study the active longitudes of sunspots and flares together.

In our previous paper (\cite{Gyenge14}, hereafter Paper I), we endeavoured to follow an as presumption-free procedure as reasonable and possible. The activity concentration has been computed for each longitudinal stripe of $10^{\circ}$ in each Carrington rotation and all these longitudinal distributions have been plotted in a time-longitude diagram. The migration of the positions of strongest concentrations in the time-longitude domain outlined a forward-backward motion of the most active longitudinal domain with respect to the Carrington frame. The half width of this active longitude was about $30^{\circ}$, a flip-flop event was identified, but the most interesting result was that the flux emergence exhibited a 1.3-year periodicity within the active longitudinal domain while this period was absent out of this domain. This raised the possible interpretation that the active regions emerging within the active longitude are anchored at the bottom of the convective zone where this period also was detected by \citet{Howe2000} and the rest of the active regions emerge from higher layers.

This interpretation offers a feasible explanation for a long enigma, the unknown connections between various space physics processes varying with a period of 1.3 years. The first report about this period has been presented by \citet{Silverman83}. They detected the 1.3-year period in century-long visual aurora observations and they admitted that the source of this behaviour was unknown. This was followed by a series of reports on this period in different space parameters: solar wind velocity \citep{Richardson94, Mursula2004}, solar wind velocity and geomagnetic activity \citep{Paularena1995, Mursula2000}, $B_{z}$ component of the interplanetary magnetic field \citep{Szabo95}, cosmic ray intensity \citep{Kudela02, Singh2012}, along with geomagnetic activity \citep{Strestik2009} refers to this period even in biological phenomena. The most recent overview and analysis of this period in various space physics processes is published by \citet{Cho2014}.

The results of \citet{Howe2000} also motivated search for this period among solar phenomena. \citet{Komm03} corroborated these results. \citet{Krivova02} found the 1.3-year period in wavelet spectra of sunspot numbers and areas without regard to its spatial relations. \citet{Obridko07} identified this period in the variation of the large-scale magnetic field by combining recent magnetograms and earlier H-alpha filament observations, the resulting time interval covered 8 solar cycles. \citet{Ruzmaikin2008} used longitudinally averaged synoptic magnetograms and detected this period too. As an important chain between the solar and interplanetary/geomagnetic phenomena, the period has also been pointed out in the time series of the flare index \citep{Ozguc2003}.

It is a common feature of the above studies that the 1.3-year period is considered as a global feature, its localisation on the solar surface is, however, not addressed. However, Paper I presented this localisation. The period can be detected in the variation of the flux emergence within the identified active longitude while it is absent elsewhere. This promises a more direct chain between the dynamics of the tachocline zone and the surface as well as that of the interplanetary space.

The aim of the present work is to specify the spatial and temporal characteristics of the flare activity in connection with the behaviour of the active longitude presented in Paper I.

\section{Databases of flares and sunspots}

The \fnurl{Debrecen Photoheliographic Data sunspot catalogue}{http://fenyi.solarobs.unideb.hu/DPD/index.html} (DPD, provided by Debrecen Heliophysical Observatory) \citep{Gyori11} is used for determining the AL. The DPD is based on white-light images of sunspot groups starting at 1974, so this study can cover 540 Carrington Rotations  (hereinafter CR), which roughly equals to five solar cycles.  The time resolution of the sample is one observation for each day. The catalogue provides us with information about the following properties of every sunspot of sunspot groups: date of observation, position of the spot and umbra and whole spot area. The estimated mean accuracy of the position data is 0.1-0.2 heliographic degrees and the estimated mean precision of the area measure is $10-20\%$. For that reason, this dataset could provide probably a very detailed sunspot survey hence it is suitable to reveal the properties of the longitudinal distribution of the sunspot groups.

We used two solar flare databases. The first dataset was taken from the \fnurl{Geostationary Operational Environmental Satellite system}{http://www.ngdc.noaa.gov/stp/satellite/goes/} (GOES) starting at 1976. The flare catalogue contains information about the date, position of the eruption, the associated active region and the classification code (measured in $W/m^2$), which can be of type A, B, C, M or X according to peak flux of the solar flare observed in the $0.1$ and $0.8$ nm wavelength range. The most energetic class consists of the X-type flares with fluxes in excess of $10^{-4} W/m^{2}$ at Earth.

Other data source was observations by the \fnurl{Reuven Ramaty High Energy Solar Spectroscopic Imager}{http://hesperia.gsfc.nasa.gov/hessidata/dbase/} (RHESSI) satellite \citep{Lin02}, containing data since 2002. The dataset currently contains more than 100,000 events in the energy band from soft X-ray to gamma ray. The flare catalogue provides the following information for each candidate: dates of explosions, durations, peak intensities, and total counts number of the X-ray photons during the flare event, energy channel of the maximal energy at which the flare is still measurable, location on the solar disc, and quality flags. Based on the positions of certain eruptions we added associated active regions by using DPD where the RHESSI flare list has not contained it. This satellite is able to observe less intense events then the GOES satellite. The RHESSI flares are mostly micro-flares of GOES class A, B, or C. The most frequent type of RHESSI flare is GOES class B.

\begin{figure}
	\centering
	\includegraphics[width=88mm]{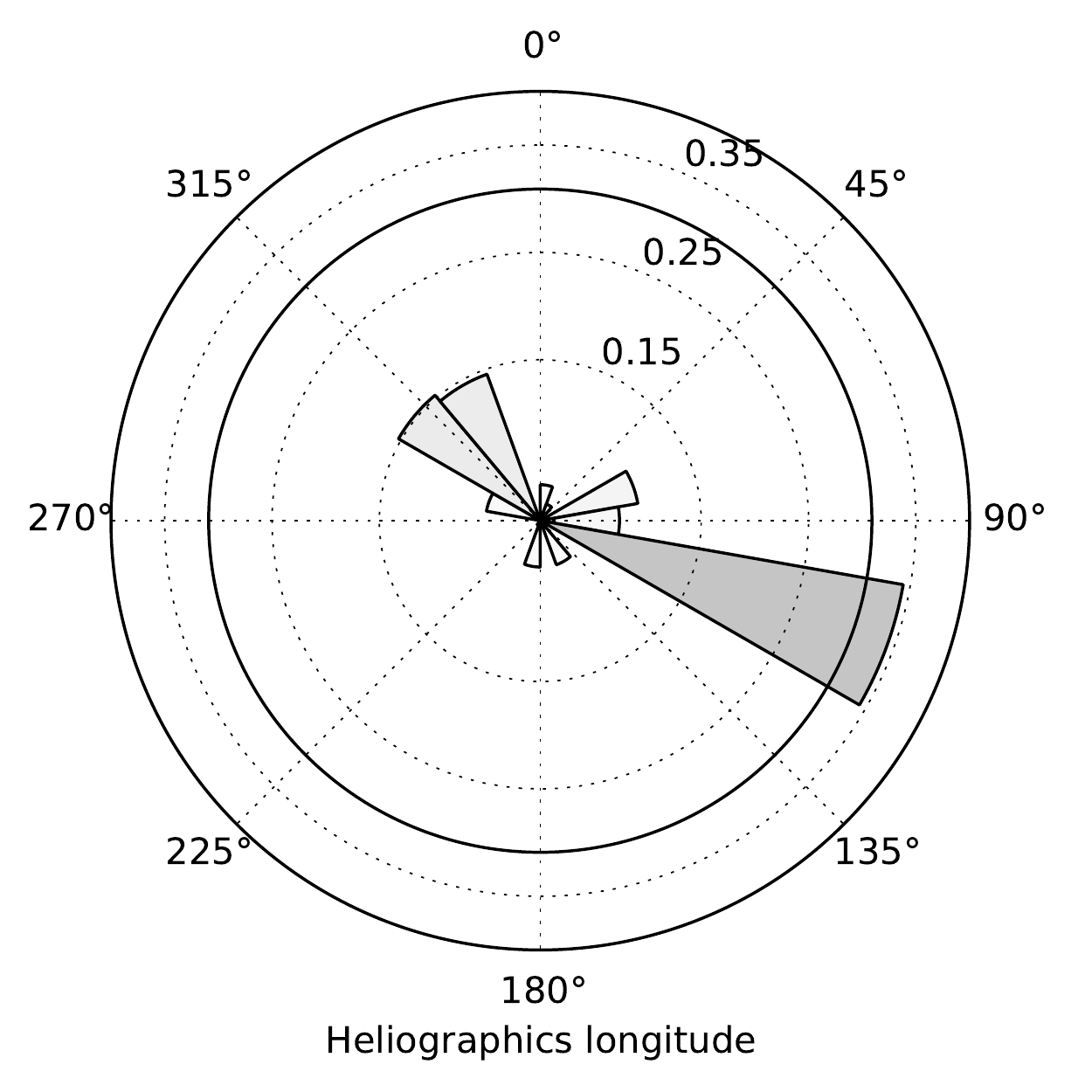}
	\caption{An example for the spatial distribution of the quantity $W$ in the CR 1718. The black line represents the 3$\sigma$ significance limit. The largest peak represents the AL. In this CR at the other side (roughly $180^{\circ}$ phase shift) the weaker and more disperse peaks correspond to the co-dominant AL.}
	\label{example}
\end{figure}

\section{The method of tracking active longitude}

The present investigation began with a similar method as Paper I. The areas and positions of all sunspot groups were considered at the time of their maximum area, the solar surface has been divided into longitudinal bins with a width of $20^{\circ}$. The areas of all groups were summed up in each bin ($ A_{i}$) and normalised by the summarised sunspot group area from the whole surface. Then, the longitudinal activity concentration is represented by the quantity:

\begin{equation}
 	W_{i,CR}=\frac{A_{i,CR}}{\sum_{j=1}^{18}A_{j,CR}}.
\end{equation}

In every Carrington rotation, the histogram of the quantity $W_{i,CR}$ was determined and the significance level estimated (set by 3$\sigma$) to filter out the unnecessary noise. The novelty of the present method is that all of the quantities $W_{i,CR}$ have been omitted in the further statistics that were lower than the previously mentioned limit. Figure \ref{example} shows an example of the histogram of the quantity $W$ in CR 1718. By filtering the data with this limit, $53\%$ of the CRs shows exactly one significant longitudinal belt meanwhile the remaining fraction of the data did not show strong in-homogeneous properties.

Many studies have adressed the existence of the co-dominant longitude that has a phase shift of $180^{\circ}$ from the dominant peak \citep{Usoskin05, Zhang07, Zhang2011a}.  However, we have found that the secondary AL is not or hardly detectable after applying a high-pass filter to sort out the insignificant peaks. For example, by setting the significance level to 2$\sigma$, only $23 \%$ of the CRs have some sort of double active longitude distribution. For that reason, the activity at the co-dominant longitude is considered weak and rare, therefore we neglect it in this study. 

\begin{figure}
	\centering
	\includegraphics[width=85mm]{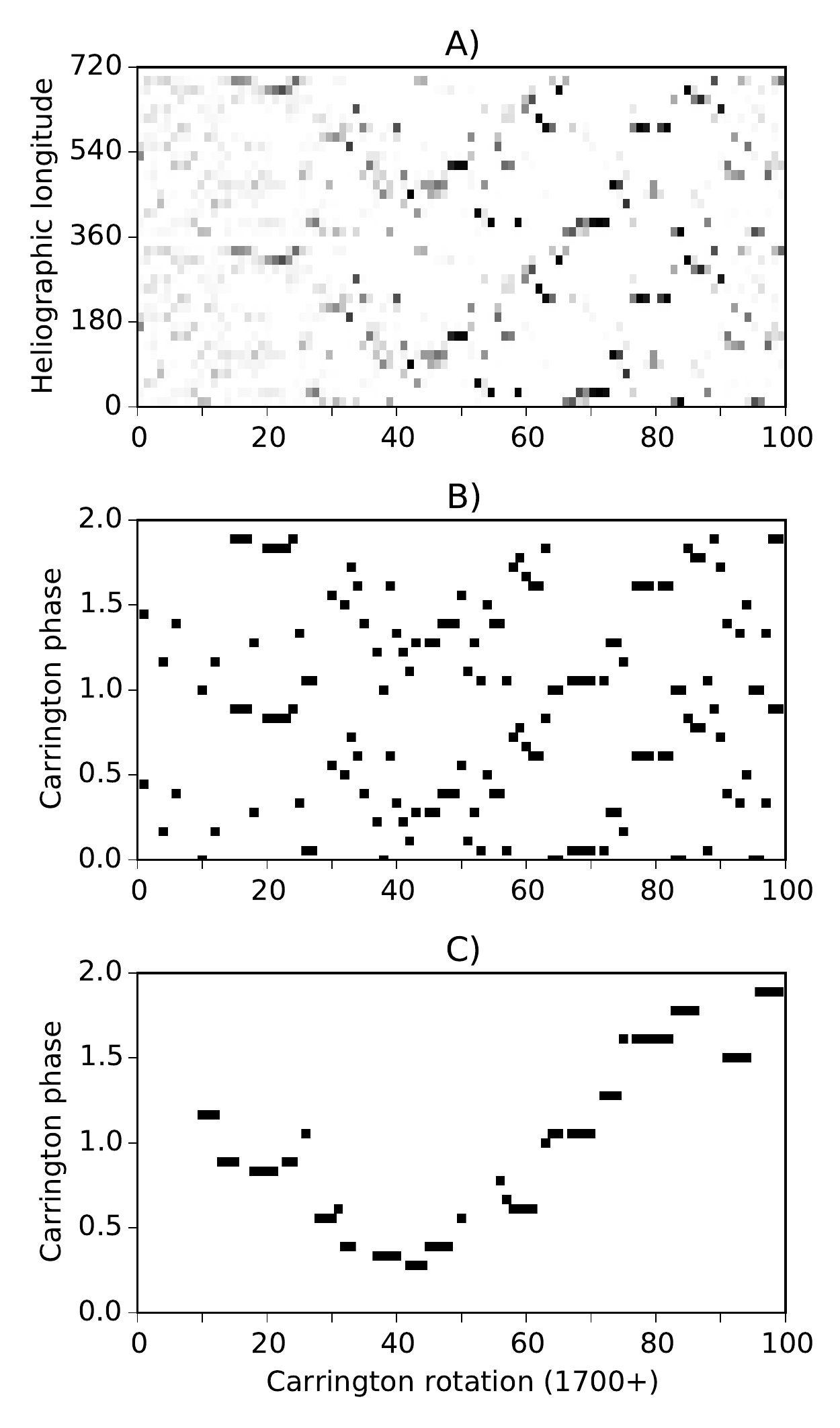}
	\caption{Different stages of the AL finding method.  Panel $A$ shows the values of parameter $W$ in longitudinal bins of $20^{\circ}$ and temporal bins of 1 CR. The darker marker means larger value of $W$. Panel $B$ shows the Carrington phase of $W$ versus time after applying a 3$\sigma$  high pass filter to omit all of the bins with insignificant value of $W$. Panel $C$ shows the final stage of the finding method of AL. The repetitions of the pattern in Carrington phases are removed and 3-CR moving average has been applied to the remaining peaks of $W$. The result shows the migration path of AL.}
	\label{method}
\end{figure}

\begin{figure*}
	\centering
	\includegraphics[width=160mm]{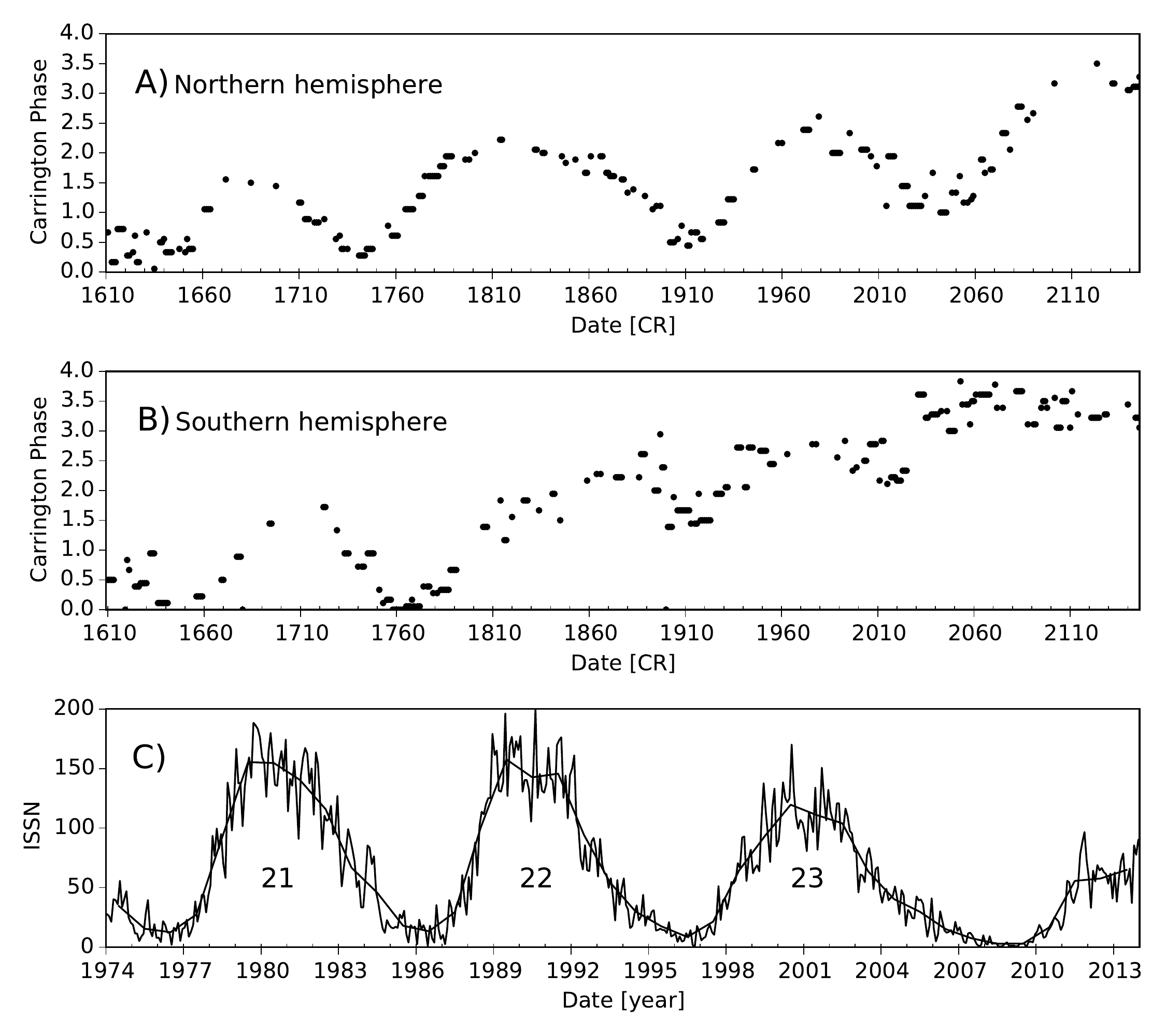}
	\caption{Activity maps for whole the DPD-era (1974-2014). Panel $A$ shows the migration of the AL on the northern hemisphere. The black markers mean the longitudinal belts of most enhanced sunspot activity (AL) in certain CRs. Panel $B$: the same as Panel $A$ but for the southern hemisphere.  Panel $C$ represents the temporal variation of the International Sunspot Number (ISSN). The numbers of the solar cycles are also marked.}
	\label{activity_map}
\end{figure*}

In Figure \ref{method}, we present an example for the steps of the identification method of the AL. There are three panels. The first panel, indicated by letter $A$, is the surface map of the parameter $W$. The horizontal $x$-axis is the time interval  between CRs 1700 and 1800. The vertical axis is the longitude in heliographic coordinate system. Panel $A$ shows the first step of our AL tracking method where all of $W_{i}$ are visible in a certain CR. In the $y$-axis the $360^{\circ}$ solar circumference has been repeated, similarly to \cite{Juckett06}, in order to follow the occasional shifts of the domains of enhanced sunspot activity in the Carrington system. The values of the parameter $W$ are represented by the colour of a certain box. $W_{i,CR}=1$ is associated by black and $W_{i,CR}=0$ is plotted by white. 

For further analysis, the Carrington longitudes will now be transformed into unit of Carrington phase:

\begin{equation}
	\psi=\lambda/360^{\circ}.
\end{equation}

Here, the value $\lambda$ is the longitudinal position in the Heliographic coordinate system, therefore the $\psi$ values must be always between 0 and 1 (which represents the entire circumference). Please note that the precision of the position of the DPD catalogue is about 0.1 heliographic degrees, for that reason the accuracy of the Carrington phase is about $\pm 2.7 \times 10^{-4}$, which is negligible.

Panel B of Figure 2 shows the temporal variation of the Carrington phase. Here, the high-pass filter has  been applied (set by 3$\sigma$) to filter out the values of parameter $W$ has been omitted which were lower than 3$\sigma$. The significant migration path is more remarkable here. 

The final step of the AL tracking method is visualised in panel C of Figure \ref{method}. We applied a moving average method for smoothing the data. The time-step is 1 CR and the width of the time window is 3 CRs. The repetitions of the AL shifted with one Carrington phase are removed.  A phase jump occurs in case of the migration of AL enters to the next solar circumference. A phase jump occurs when 
\begin{equation}
	\delta\phi > \delta\phi^{*},
\end{equation}
where the parameter $\delta\phi$ and $\delta\phi^{*}$ defined as;

\begin{equation}
	\begin{array}{rcl}
		&\delta\phi & =  |\psi _{CR+1}-\psi _{CR}| \\
		&\delta\phi^{*} & = |(\psi _{CR+1}+\gamma)-\psi _{CR}|, 
	\end{array}
\end{equation}
and $\gamma$ is the phase shift which could be $1$ or $-1$.

For instance, a phase jump can be seen in Figure 2 in CR 1765 when the curve is crossing from Phase 0-1 to Phase 1-2. The shape of the migration path may depend on how the phase jumps are handled. Although the long-term trends of the migration path may not be always well determined because of the phase jumps and the scatter of the points around it, this may hardly affect the statistical results based on the distances measured from the actual AL values. Thus, the distances from AL are used in the next sections.

\section{Tracking active longitude in the DPD era}

The AL tracking method was applied to the entire DPD-era, containing 540 CRs from  CR 1610 to CR 2150. The indicated time interval consists of 3 full (namely Cycles 21, 22 and 23) and 2 partially visible (namely Cycles 20 and 24) solar cycles. The two hemispheres are treated separately. Sunspot groups were selected within a longitudinal distance of $\pm 75^{\circ}$ from the central meridian in order to avoid the less accurate area and position estimates at the limb. In Figure \ref{activity_map}, we plot the temporal variation of the Carrington phase for the northern hemisphere (Panel A) and the southern hemisphere (Panel B). Panel C shows the temporal variation of the International Sunspot Number provided by the \fnurl{SIDC-team}{http://www.sidc.be/silso/datafiles}.

The migration of AL in the northern hemisphere is clearly recognisable. The shape of the migration seems like a set of parabola with prograde and retrograde phases. On the prograde phase the angular velocity of the AL is faster than the Carrington rotation and the retrograde phase, on the contrary; it shows a slower angular velocity. The maxima of parabola-like curves roughly correspond to the maxima of solar cycles, but the beginnings and ends of the pro and retrograde phases do not correspond to the minima of the solar cycles. In Cycles 21 and 23, the retrograde and prograde parts of the parabola are shorter and the migration shape of the AL of the Cycle 22 clearly shows a longer period in time than the period of the solar cycles. On the southern hemisphere, similar parabola shape migration pattern has been shown but the times of minima and times of maxima differ from those of Panel A and they are sometimes less expressed.

The migration of the longitude of enhanced sunspot activity has been found in numerous earlier studies. First detection of some kind of pattern of migrating AL of sunspot groups was reported by \cite{Berdyugina03}. \cite{Juckett06} found similar returning phase in the time interval between CRs 1720--1750 by using low degree surface spherical harmonics. The retrograde phase was also observed on a restricted interval by \cite{Bumba2000} for Cycle 22.

In Paper I, we concluded that the migration paths of ALs did not correspond sharply to the shape of the 11-year cycle. This statement was based on two parabolic patterns of AL, one in the northern hemisphere in Cycle 22 and one in the southern hemisphere in Cycle 24. The present extension of the study confirms this statement for the whole interval. Based on these longer time series of migration patterns, it is clearer that the minima of the patterns mostly do not coincide with those of the solar cycles. In addition, the two hemispheres are able to produce different migration patterns with different locations of minima and maxima.

For this reason, differential rotation can hardly be involved in the migration of AL. \cite{Balthasar07} did not find signatures of differential rotation either, similarly to the above-mentioned works of \cite{Juckett06, Juckett07} . Furthermore, note that the AL seems not always identifiable.  $47\%$ of the data did not show significantly enhanced longitudinal activity in the case of $3\sigma$ significance limit.

\section{Spatial distribution of solar flare occurrence}

\begin{figure}
	\centering
	\includegraphics[width=87mm]{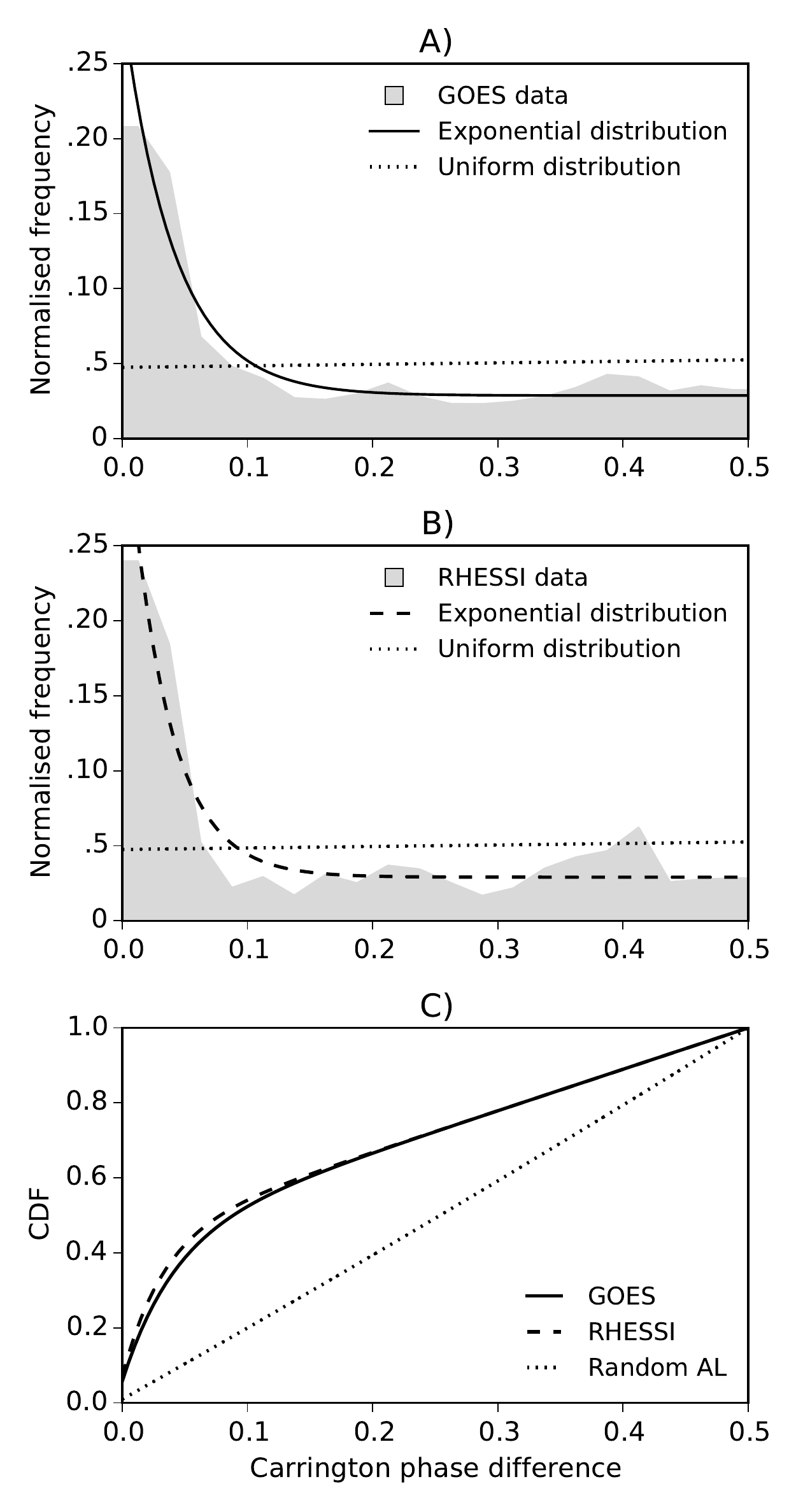}
	\caption{The spatial distributions of the parameter $\delta\psi$. Panel $A$ shows the spatial distribution of GOES flares with respect to the AL, represented by the grey area. The solid line is the fitted exponential model distribution and the dotted line means the uniform distribution based on random AL. Panel $B$ represents the same statistics but the RHESSI flare list has been used. The dashed line is the fitted exponential model distribution and the dotted line means the uniform distribution based on random AL. The last figure (Panel $C$) shows the cumulative distribution of the fitted model distributions.}
	\label{stat}
\end{figure}

In this section, we investigate the relationship between the position of solar flares (the data are taken by GOES and RHESSI satellite) and the longitude of enhanced sunspot activity (AL).  For that reason a new parameter is defined which is the shortest longitudinal distance between the position of the AL and the position of certain event. The phase difference is defined by:

\begin{equation}
	\delta\psi=\left|AL_{CR}-L_{CR}\right|.
\end{equation}

Here, $L_{CR}$ is the longitudinal position of the RHESSI and GOES observations in Carrington phases. $AL_{CR}$ represents the position of the AL defined in Section 4. The parameter $\delta\psi$ is reduced by a unit phase if it is larger than $0.5$ because the distance cannot be larger than 180 degrees.  $\delta\psi=0$ means that the flare is located at the AL; $\delta\psi=0.5$ means that the flare is located on the other side of the Sun opposite the AL.

Figure \ref{stat} contains three plots where panels $A$ and $B$ show the spatial distribution of the RHESSI and GOES flares with respect to the AL (the position of the AL corresponds to $\delta\psi=0$). The grey area represents the frequency distribution normalised by the sample number. On the plots, there is only one significant peak for each statistics and after that a long plateau with insignificant local peaks. The significance limit was defined by the standard deviation of the sample ($\sigma_{GOES}=9.8$ and $\sigma_{RHESSI}=10.5$). The distributions show that most of the solar flares are concentrated in a relatively narrow longitudinal zone. In Paper I, we found that AL of sunspot groups is as narrow as $20-30^{\circ}$. Here, the spatial distribution of solar flares shows similar behaviour. To estimate the quantitative properties of this distribution, different models have been tested and the exponential distribution has provided a good enough result for each dataset. The best exponential models are indicated by solid line in panel $A$ and by dashed line in panel $B$. 

 We defined control groups where the position of the AL generated by random numbers. This test was inspired by \cite{Pelt05} who criticised the AL identification method of \cite{Berdyugina03}. \cite{Pelt05} reconstructed the distribution of the AL with random sunspot longitudes. For that reason, we repeated the statistical study by using the identification the AL based on random longitudinal sunspot positions. The result of the control group does not show significant peaks, thus it can be represented by dotted line of homogeneous distribution in Figure \ref{stat}. This means that our AL identification method does not cause false significant peaks, which would affect the results of the study of in-homogeneous longitudinal properties.

Panel $C$ of Figure \ref{stat} shows the cumulative distribution of the spatial distributions. The solid and dashed lines (based on GOES and RHESSI observations) have a steep increasing phase between the values of $0$ and $0.1$ after the functions show a constantly less steep increasing trend. These models allow to estimate that the most of solar flares (around $60\%$) occur in the longitudinal zone of $\pm 36^{\circ}$ around the position of AL. The dotted line is the cumulative distribution of the uniform distribution based on random longitudinal positions. In this distribution the mentioned longitudinal domain would only contain $20\%$ of the solar flares. This means that the AL plays a significant role in the spatial distribution of flare occurrences.

\section{Temporal properties of solar flares near active longitude}

\begin{figure}
	\centering
	\includegraphics[width=87mm]{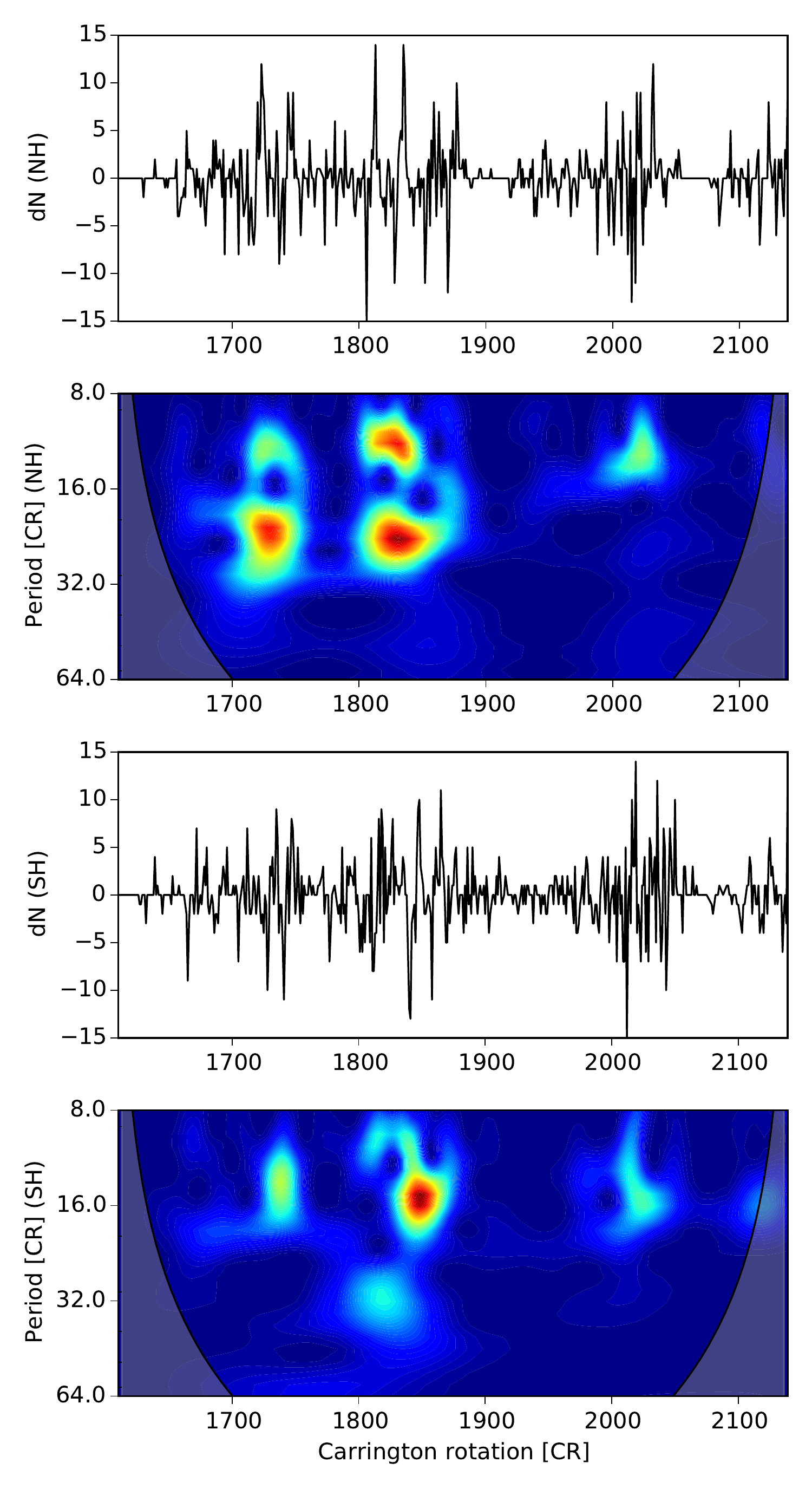}
	\caption{The temporal distribution of the solar flares near to the dominant AL. The smoothed solid black line of the first and the third subplots mean the time variation of the GOES solar flare variance. Below those plots, continuous wavelet transforms of the time-series are presented.  The first and the second subplots contain data from the northern hemisphere (NH) and the last two subplots show the southern hemisphere (SH). The bottom axis is time in CR for each subplot. The red contours enclose regions of greater than $95\%$ confidence. The grey faded regions on either end of the figures indicate the cone of influence (COI) where the edge effect becomes significant for that reason all of the found peaks under the COI has been omitted.}
	\label{wavelet}
\end{figure}

\begin{figure}
	\centering
	\includegraphics[width=87mm]{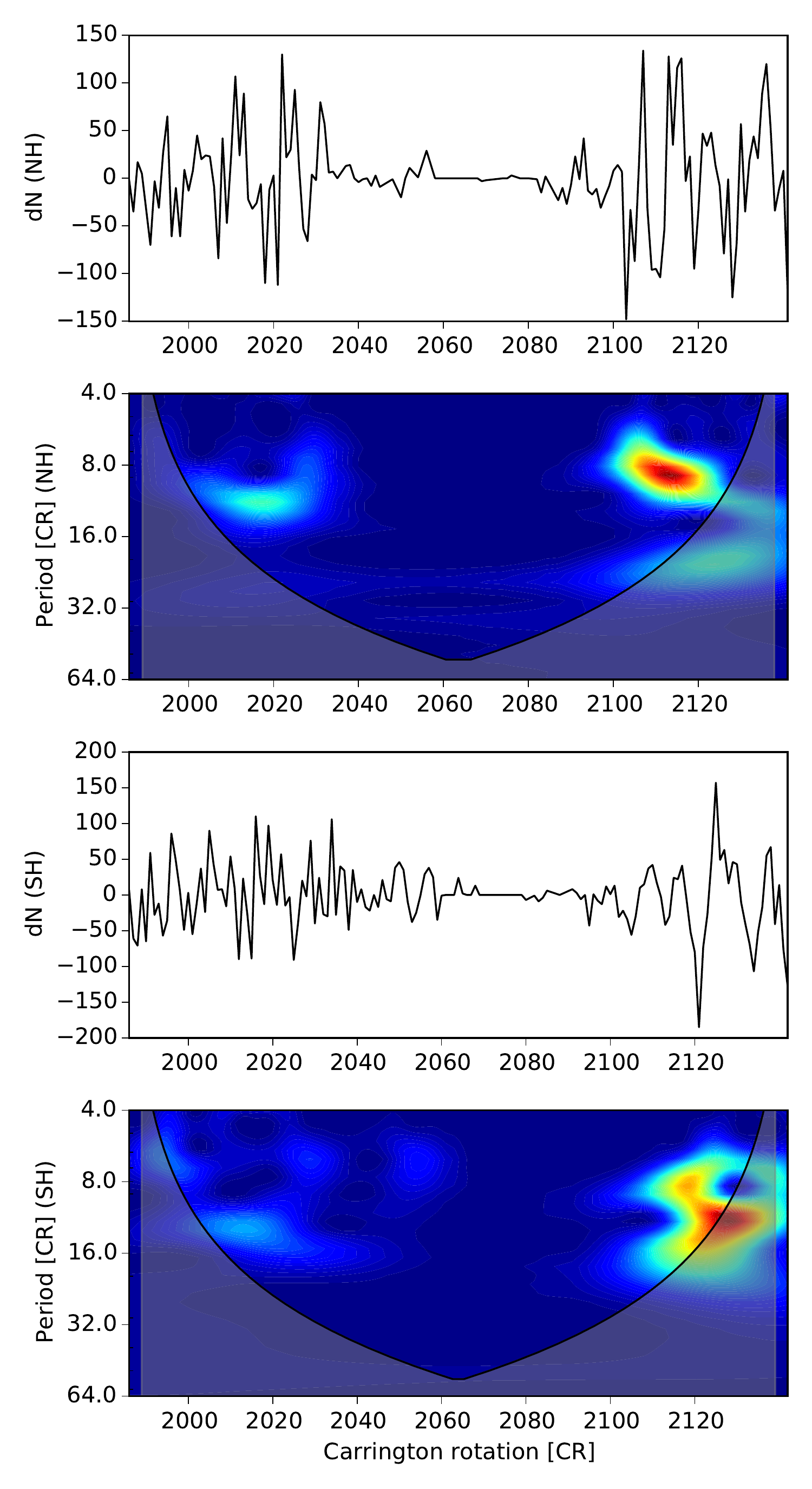}
	\caption{The same statistics as the Figure 6 but the data was taken from the RHESSI catalogue. The bottom axis is time in CR. The indicated period covers the whole RHESSI database from 2002. The meaning of the colour codes are the same, the red contours encloses regions of greater than $95\%$ confidence. The grey faded regions on either end of the figures indicate the cone of influence (COI) where the edge effect become significant for that reason all of the found peaks under the COI has been omitted.}
	\label{wavelet2}
\end{figure}

The aim of this section is to reveal the temporal properties of the occurrence of solar flares located near to the AL. We showed in the previous section that most of the solar flares detected by the RHESSI and GOES satellite concentrate in a longitudinal strip of $0.1$ Carrington phase width around the AL. Here, we focus only on this specific region. We omit all of the solar flares which occur further than the $0.1$ Carrington phase. This spatial limit corresponds to a neighbourhood of $\pm 36^{\circ}$ of the dominant AL in the Carrington coordinate system. 

In every CR, we summed-up the number of all GOES and RHESSI flares and the two hemispheres are distinguished. The parameter $dN$ is defined for this statistics:

\begin{equation}
	dN=N_{CR}-N_{CR+1}.
\end{equation}

In Figure \ref{wavelet}, the results of the statistics are plotted for data provided by the GOES satellite. The first (northern hemisphere) and third (southern hemisphere) panels are the temporal variations of the parameter $dN$. The time series is smoothed by the moving average method with 4-CR smoothing window and 1-CR step. Below each plot, the wavelet spectra or the parameter $dN$ are presented. All of the flares are taken into account for both statistics.

In the GOES-era four significant periods (confidence level set by $95\%$) and several clearly visible, but not significant, peaks are found on both hemispheres. In the northern hemisphere (NH) during Cycle 21 (CR1710 - CR1750) there is a fluctuation with 18-24 CR period (indicated by the entire range of the red contour). The same period is found in Cycle 22 (CR1810-CR1750). This fluctuation corresponds to 1.3-1.8 years. There is one more significant fluctuation in Cycle 22 with about 11-13 CRs, which corresponds to 0.8-1 year fluctuation period. The same period is also found in Cycles 21 and 23 but the power of the peaks are below the $95\%$ confidence level. In the southern hemisphere (SH), only one significant peak is found in Cycle 22. The period is 14-16 CR (1 - 1.2 years). This period is also visible, below the CI $95\%$, in Cycles 21 and 23.

\begin{table}
	\center
	\caption{Significant periods (CI $95\%$) obtained by wavelet analysis based on the time series of the parameter $dN$.}
	\begin{tabular}{lllll}
 		\hline
			Time [CR] & Period [CR] & Period [yr] & Hemi & Source\\
		\hline
			1710 $-$ 1750 & 18 $-$ 24 & 1.3 $-$ 1.8 & NH & GOES \\
			1810 $-$ 1850 & 18 $-$ 24 & 1.3 $-$ 1.8 & NH & GOES \\
			1810 $-$ 1850 & 11 $-$ 13 & 0.8 $-$ 1.0 & NH & GOES \\
			1850 $-$ 1870 & 14 $-$ 16 & 1.0 $-$ 1.2 & SH & GOES \\
			2105 $-$ 2120 & 8 $-$ 10 & 0.6 $-$ 0.7 & NH & RHESSI \\
			2115 $-$ 2120 & 8 $-$ 10 & 0.6 $-$ 0.7 & SH & RHESSI \\
		\hline
	\end{tabular}
	\label{table}
\end{table}

Figure \ref{wavelet2} shows a similar statistics for the data taken from the RHESSI catalogue. The RHESSI-era starts in 2002 hence only Cycles 23 and 24 are covered. In the northern hemisphere, one significant period can be seen, corresponding to 0.6-0.7 years in Cycle 24 (2105CR -- 2120CR). The periods of 0.8 and 1.3 years are also weakly visible in Cycles 23 and 24 but the powers of the fluctuations are, again, under the $95\%$ confidence level. In the southern hemisphere, only one peak is visible in Cycle 24. The period is around 1.3 years but, unfortunately, this peak is within the COI. The last two rows of Table 1 only contain RHESSI data about a period of 0.6--0.7 years but this period was not detectable in the GOES measurements. Because of this contradiction in this time interval the suspected connection of surface activity with the dynamics of the tachocline layer probably does not work.  The found significant periods are summarised in Table \ref{table}.

\section{Discussion and Conclusion}

\cite{Zhang08} found that the active longitudes with half width of $20-30^{\circ}$ contain 80\% of C-flares during the solar minimum and X-flares during solar maximum. The active longitude was defined by the dynamic reference frame, introduced by \cite{Usoskin05}. Hence, their study assumes two detectable and equally strong active longitudes, separated by $180^{\circ}$. In our study, however, the secondary active longitude was found to be weak in comparison with the primary one. Our results imply that only the primary belt contains 60\% of all solar flares. The entire width of the belt is $\pm 36^{\circ}$. Both values correspond well to the results, investigated by \cite{Huang13} who calculated the ratio between the number of flaring active regions and total number of active regions. The active region that is near the active longitude is prone to erupt. The active region that is far away from the active longitude produces solar flares with little probability. 

Numerous earlier studies reported periods between 1 and 2 years in a number of heliospheric parameters such as the interplanetary magnetic field \citep{Vilppola01}, the solar wind speed \citep{Mursula03}, the cosmic rays, the geomagnetic activity \citep{Paularena1995}, coronal index and the 10.7-cm solar flux data \citep{Forgacs2007}. \cite{Mursula03} found mid-term quasi-periodicities based on the 'aa' index of global geomagnetic activity. Periods found correspond to $1.2-1.4$ years and a slightly longer period of about $1.5-1.7$ years.\citep{Cho2014} found significant signals of 1.3 years between 1987 and 1995 in the solar wind speed, IMF $B_{z}$, geomagnetic aa index and ap index as well as in the tachocline layer $\Delta \Omega$ after 1997.

The 1.3-year period of flare emergence in the active longitude found here corresponds well with our previously found 1.3-year period of spot emergence. It has been shown in Paper I, that this period can only be detected in a narrow belt of $30^{\circ}$ along the parabola-shaped path of AL but it is absent, or overwhelmed, in the entire material. The presence of the $\sim$ 1.3-year period within the active belt allows the conjecture that the active longitudes may be connected to a source region close to the tachocline zone. This implication is based on the results by \cite{Howe2000}, who found the same period of radial torsional oscillation at the tachocline zone during the period between 1995 and 2000 and by \cite{Bigazzi04}, who argued that the non-axisymmetry of the solar dynamo could only remain at the bottom of the convective zone.

It may be interesting to investigate whether the 1.3-year fluctuations in the tachocline zone and the interplanetary phenomena may be connected. It could be possible because the oscillation of the tachocline zone, the flux emergence of the active longitude and the flare occurrence within the active longitude show similar temporal properties. However, to reveal this connection is not an easy task because on the one hand the time intervals of their observations are not continuously overlapped. On the other hand this fluctuation is not always present in any of these regions.

For the period 1995 and 2000, Paper I also presented a clear signature of the 1.3 year fluctuation in the magnetic flux emergence within the AL, the targeted time interval has a short overlap with that studied by \cite{Howe2000}. The first two intervals in Table I are also parts of this interval and they also contain the 1.3-year periodicity of flaring activity. Thus, in spite of the incomplete observational coverage it can be assumed that the excitation of the fluctuation at the tachocline layer may be connected with the dynamics of flux emergence in the AL and with the fluctuation of the flaring activity and interplanetary impact. The temporal variation of the number of solar flares could be related with the flux emergence based on the simple idea that, the probability of a flare occurrence is higher if there is an increased number of the active regions. However, there is an uncertainty as to why the different hemispheres could produce different fluctuations. This question definitely needs further study.

\cite{Howe2011} concluded that the significance of the 1.3-year fluctuation became less significant between 2000 and 2010 and at the end of the indicated time period the 1.3-year fluctuation have not found. There are no results published after 2010. However, after 2000 we have also not detected this period in the flaring activity or in the temporal properties of the sunspot flux emergence (Paper I).

\section{Summary}

Many flare prediction models employ the properties of active regions, such as morphological information, area or the magnetic field of sunspot groups. The spatial distribution of active regions has not been used widely, such as longitudinal distribution. If we assume the most flare-productive active regions tend to be located in or close to the active longitudinal belt then, this may allow to predict the geo-effective position of the domain of enhanced flaring probability for a couple of month or years ahead. In this paper, we studied the spatial and temporal properties of solar flares based on observations of the GOES and RHESSI satellites.

We have found that there is a narrow longitudinal belt of considerably enhanced sunspot activity migrating in the time-longitude domain. The migration path is similar to of a series of parabolic shape curves. A parabola-like section of the migration path does not follow the solar cycle and it has fluctuating temporal properties. The co-dominant longitudinal belt, phase shifted by $180^{\circ}$, is relatively weak in comparison to the main active longitude. The main active longitude is not always detectable but the migration path of the activity is recognisable based on the DPD sunspot data in time interval between 1974 and 2014.

Our results shows that the main active longitude plays a crucial role in the global position of solar flare occurrence. The most flare active groups appear in the $\pm 36^{\circ}$ width of the active longitude. The temporal variation of the number of solar flare shows fluctuations 0.8, 1.3 and 1.8 years within the active longitudinal belt. This temporal distribution may also provide an improved flare forecasting opportunity.

\section*{Acknowledgments}

NG would like to thank for the invitation, support and hospitality received from Kalevi Mursula (Department of Physical Sciences, University of Oulu, Finland). The research leading to these results has received funding from the European Community's Seventh Framework Programme (FP7/2007-2013) under grant agreement eHEROES (project No. 284461). This research was made use of SunPy, an open-source and free community-developed Python solar data analysis package \citep{Mumford}. Finally, NG thanks for the support received from the University of Sheffield.

\end{document}